# Magnetodielectric effect of $Bi_6Fe_2Ti_3O_{18}$ film under an ultra-low magnetic field

J. Lu [1], L.J. Qiao [1], X.Q. Ma [2] and W.Y. Chu [1]

[1] *Department of Materials Physics, University of Science and Technology Beijing, Beijing 100083, People's Republic of China*

[2] *Department of Physics, University of Science and Technology Beijing, Beijing 100083, People's Republic of China*

E-mail: lqiao@ustb.edu.cn (L.J. Qiao)

**Abstract**

Good quality and fine grain $Bi_6Fe_2Ti_3O_{18}$ magnetic ferroelectric films with single-phase layered perovskite structure have been successfully prepared via metal organic decomposition (MOD) method. Results of low-temperature magnetocapacitance measurements reveal that an ultra-low magnetic field of 10 Oe can produce a nontrivial magnetodielectric (MD) response in zero-field-cooling condition, and the relative variation of dielectric constants in magnetic field is positive, i.e., $[\varepsilon_r(H)-\varepsilon_r(0)]/\varepsilon_r(0)=0.05$, when T<55K, but negative with a maximum of $[\varepsilon_r(H)-\varepsilon_r(0)]/\varepsilon_r(0)=-0.14$ when 55K<T<190K. The magnetodielectric effect appears a sign change at 55K, which is due to transition from antiferromagnetic to weak ferromagnetic; and vanishes abruptly around 190K, which is thought to be associated with order-disorder transition of iron ion at B site of perovskite structures. The ultra-low-field magnetodielectric behaviour of $Bi_6Fe_2Ti_3O_{18}$ film has been discussed in the light of quasi-two-dimension unique nature of local spin order in ferroelectric film. Our results allow expectation on low-cost applications of detectors and switches for extremely weak magnetic fields in a wide temperature range 55K-190K.

## 1. Introduction

Magnetoelectric materials, which possess at least magnetic and ferroelectric ordering in certain range of temperature and therefore promise applications on smarter multifunctional devices, have entered a spotlight of scientific community [1]. Quite a few researchers have been interested in combination of high permittivity of ferroelectrics with magnetic degree of freedom of manipulation, which has potential applications on non-contact control of dielectricity on insulating devices [1]. Magnetodielectric (MD) effect means the relative variance of dielectric constants after and before applying magnetic fields, i.e., MD=[$\varepsilon_r(H)-\varepsilon_r(0)$]/$\varepsilon_r(0)$ [2]. Significant MD effect, such as giant magnetocapacitance with peak MD=5 and colossal magnetocapacitance also with peak MD=5, has been discovered recently in $DyMnO_3$[2] and $CdCr_2S_4$[3] single crystal, respectively. Rather high magnetic fields with H>1x10$^4$Oe are usually inevitably needed for the high MD effects [2]. Remarkably low fields, e.g. in the magnitude of 10$^3$Oe, induced maximum MD=0.03 in bulks of tebium iron garnet [4]. Solid state materials other than bulk have been attempted and exhibit sensible MD responses even under quite a low magnetic field of 50 Oe [5], but one may find that their origins are non-intrinsic or indirect, such as magnetoelastic or magnetotransport effects under a small length scale, rather than coupling directly between magnetic field and polarizability [5]. Since sensible intrinsic magnetodielectric effects under ultra-low magnetic fields have not yet reported, hence, our efforts have been devoted to the study on magnetodielectric behaviours of thin films of appropriate mutiferroic materials under an ultra-low magnetic field.

Bismuth-layered structure ferroelectric family has a general formula $Bi_2M_{n-1}R_nO_{3n+3}$ where M=Bi or rare earths and R=Fe/Ti [7]. In this paper, $Bi_6Fe_2Ti_3O_{18}$ magnetic ferroelectric film is used. Bismuth layered magnetoelectrics doped with iron has been proved novel in multiferroic output [8], since bismuth layered structures suggest excellent ferroelectricity [9] and magnetic ordering easily forms after introducing iron ion partially on pseudoperovskite B sites [10]. So it was decided to study coexistence of ferroelectrics and magnetism in bismuth layered titanate via iron substitution. To synthesize thin films, metal organic decomposition (MOD) method was chosen because it has couples of merits to deposit oxide coatings, i.e., lower temperature, non-vacuum, and especially low cost and extensive tailorability [11].

## 2. Experimental

As for wet chemistry method, precursor is evidently crucial to the structure of final film [11]. In order to synthesize the precursor of the $Bi_6Fe_2Ti_3O_{18}$ film, hydrated bismuth nitrate with 10mol% Bi excess to compensate its volatility in high temperature when annealing, and stoichiometrical hydrated iron nitrate were first put into 2-methoxyethanol, with glacial acetic acid as organic ligand agent and ethylene glycol to crosslink metal-organic molecules [11], and then stoichiometrical tetrabutyl titanate was put into this solution. It is noteworthy to mention that the molar weights of acetic acid and ethylene glycol are equal to that of nitric ion and half of aggregate metal molar weight, respectively. The specific steps are schematically drawn in Figure 1. Emphasis on aging of mixture is worthwhile because of its nontrivial effect on micro morphology and electric properties of films [13].

For experimental convenience, N-type Si wafer with (111) orientation and a thickness of 500μm was cut into pieces with size around 1.2x1.2cm$^2$. After series of cleaning and oxidation procedures [14], a layer of amorphous platinum with thickness of about 100nm was sputtered onto $Si/SiO_2$ substrate, then the Pt bottom electrode was crystallized at 400 ℃ for 1 hour in ambient atmosphere before deposition of metal-organic sol.

The samples were obtained by spin coating and rapid thermal processing [15]. The spinning parameters were set at 4000rpm for 30s. Wet films were subsequently placed onto hot plates with temperatures at 160-180℃ for 1 minute to vaporize solvent and other volatiles, and at 400-500℃ for 10 minutes to nucleate Bi-layered structure, respectively [16]. Usually, we repeated the above steps 4 times so as to obtain the final multilayer film with a thickness around hundreds of nanometers. Ultimately, the multilayer film were subject to furnace at 650-750℃ for 60 minutes so as to grow grain sizes [16]. Before electric characterizations, each sample was also needed to be deposited a thin platinum layer (30-100nm) as top electrode via mask technology, where the round electrode areas were confined in the range of 0.04-1mm$^2$.

Optical microscope and scanning electronic microscope (SEM) were used to diagnose morphology. Thicknesses were also estimated by means of SEM. X-ray diffraction method using Cu Kα radiation acted as our major resorts to reveal crystallogram. Our installation of facility for dielectric measurements was according to Sawyer-Tower circuit [17] where signal applied on films was 1kHz sine wave and a 10nF standard capacitor in series with the sample was used to measure charges. In addition, during testing the samples were placed in Cryogen Free Superconducting Magnet & VTI System produced by CRYOGENIC, where the direction of magnetic field was perpendicular out of the films.

**3. Results and discussion**

The $Bi_6Fe_2Ti_3O_{18}$ films obtained have good quality, as shown in Figure 2. The whole surface with size of 13x13mm$^2$ is smooth, as shown in Figure 2a and Figure 2b. Figure 2c indicates that the film with grain size of about 150nm is crack-free and pinhole-free.

The structure of $Bi_6Fe_2Ti_3O_{18}$ film is revealed with comparison to powder derived from corresponding sol and with the same treatment condition as that of films, as shown in Figure 3. Figure 3a is XRD pattern of the $Bi_6Fe_2Ti_3O_{18}$ film and Figure 3b is that of the powder. Figure 3c is XRD pattern of $Bi_4Ti_3O_{12}$ powder which is cited from JCPDS file 35-0795 and suggests its nature of single phase. Comparing Figure 3a to Figure 3b and Figure 3c, it is understandable that the $Bi_6Fe_2Ti_3O_{18}$ film possesses single-phase structure similar to $Bi_4Ti_3O_{12}$ since the reflection positions agree well with that of powder of $Bi_6Fe_2Ti_3O_{18}$ and $Bi_4Ti_3O_{12}$. Difference in intensities may be attributed to the disorder at the Fe/Ti site [10].

The results of physical property tests have been presented in Figure 4. Figure 4a is dielectric constants during warming and cooling without magnetic field, i.e. $\varepsilon_r(0)$, and that during warming in magnetic field of 10 Oe after zero-field cooling, i.e. $\varepsilon_r(H)$. Judging from Figure 4a, we find no obvious changes occur during the warming and cooling processes without magnetic field, however, $\varepsilon_r(H)$ during warming in magnetic field of 10 Oe is evidently different from $\varepsilon_r(0)$ during warming without magnetic field. It is worth noting that the magnitude of presented dielectric constants without magnetic field is comparable with the bulk bismuth titanate in single crystals and polycrystals [18]. Therefore it is reasonable to exclude the extrinsic contributions from the Schottky barriers at electrical contacts [19].

Figure 4b is the relative variance of dielectric constant, i.e. MD=$[\varepsilon_r(H)-\varepsilon_r(0)]/\varepsilon_r(0)$ which is obtained based on Figure 4a, versus warming temperature. There are two major features in our MD findings, as shown in Figure 4b, i.e. a sign change of magnetodielectric responses takes place at 55K; and an abrupt attenuation of magnetodielectric effects exists around 190K. Experiments showed that there was a magnetic transition from antiferromagnetism (AFM) to weak ferromagnetism at 65K for bulk $Bi_6Fe_2Ti_3O_{18}$ materials [8]. We have scanned the magnetodielectric properties around 55K several times in a back and forth protocol and this transition reappeared with high reproducibility. Therefore, a sign change of MD at 55K should be due to the magnetic transition from antiferromagnetism to ferromagnetism, and the difference of this transition point between film at 55K and bulk at 65K may be explained by the constrain effect from the substrate [20].

It should be not unexpected that a second conversion takes place around 190K in Figure 4b and we make some detailed discussions here. Bearing in mind that room-temperature disorderliness exists at B site in iron modified bismuth titanate, where iron ions partially substitute $Ti^{4+}$ [10]. On the other hand the presence of local orderliness with a kind of short range interaction is understandable when temperature gets sufficiently low. It is true that for the systems of iron substituted perovskite order-disorder transition can be observed in $\chi$–T curves during cooling without magnetic field [21]. For $Bi_6Fe_2Ti_3O_{18}$ bulk materials, an order-disorder transition of iron ion at perovskite B site occurred around 170-190K and produced a peak around 170-190K in plot of magnetic-induced-electric-polarization output versus temperature [8]. Hemberger et al. also suggested that colossal magnetodielectric effects of $CdCr_2S_4$, found at low temperature, raise from coupling between relaxor ferroelectrics and magnetic fields where long-range order loses and local order differs substantially from global symmetry [3]. For $Bi_6Fe_2Ti_3O_{18}$ films, similar to the bulk materials, there is also an order-disorder transition of iron ion at perovskite B site around 170-190K. As a consequence, abrupt diminution of magnetodielectric effect around 190K at Figure 4b should be due to the order-disorder transition of iron ion at perovskite B site of $Bi_6Fe_2Ti_3O_{18}$ films.

Intriguing questions might be raised on the sign of magnetodielectric effects. Conventionally negative magnetodielectric responses happen because magnetic fields suppress the excitation from a singly occupied state to another one [22]. The positive magnetodielectric responses below 55K in Figure 4b are attributable to that magnetic fields play a role of preventing low-temperature reduction in dielectric constants when nearest-neighbor spin correlation increases for antiferromagnetic structures [23].

Magnetodielectric effects with maximum MD=0.03 in low magnetic fields of 1-2kOe were interpreted by magnetic domain occupancy variation, specifically, domain reorientation accompanied by a huge magnetostriction in $Tb_3Fe_5O_{12}$ [4]. Since in bulk materials including $Bi_6Fe_2Ti_3O_{18}$, so far no observation of MD effect under ultra-low magnetic field with the magnitude of 10 Oe has been reported, therefore, the results presented in this letter strongly suggest the uniqueness of thin films can be regarded as reasons for ultra-low-field magnetodielectric behaviors, as well as high sensitivity of magnetodielectric from compositions and structures [24]. Thin film can be considered as a quasi-two-dimension system and the quantum interference becomes important, where the spin-orbit interaction has only the z-component and magnetic scattering could set in dominantly under very low magnetic fields [25]. When weak magnetic fields are applied perpendicular to the surface of films, time-reversal invariance breaks because external magnetic fields impose a valid perturbation on phase coherence of spin, to produce large magnetoelectronic responses [26]. Magnetoresistance experiments showed that extremely weak magnetic fields below 100 Oe induced significant change of resistance in semiconductor films with thickness from 0.1μm to 1μm [26].  So it is reasonable to contemplate magnetodielectric behaviour with maximum MD=-0.14 under an ultra low magnetic field of 10 Oe here in the $Bi_6Fe_2Ti_3O_{18}$ film originates from interaction between weak magnetic field and phase coherence of local spin possessed by Fe. Theoretical study has revealed that in perovskite-structure $PbFe_{0.5}Nb_{0.5}O_3$ atom Fe has contribution to the ferroelectricity [28], which is also a collective behaviour caused by coherent shift of center atoms for perovskite structures. Subsequently, in this case, polarization of $Bi_6Fe_2Ti_3O_{18}$ film under a weak magnetic field of 10 Oe was changed as a consequence of the interplaying between spin coherence and ferroelectric activity at Fe sites. Detailed work is now ongoing and current authors call for further comments on quantitative researches.

**4. Summary**

Good quality and fine grain $Bi_6Fe_2Ti_3O_{18}$ films can be obtained by metal organic decomposition processes. The film reveals an intrinsic zero-field-cooling magnetodielectric effect under an ultra low magnetic field of 10 Oe. A sign change of MD from positive to negative occurs at 55K and is due to transition from antiferromagnetism to weak ferromagnetism. The magnetodielectric effect vanishes abruptly around 190K which is due to order-disorder transition of iron ion at perovskite B site. Ultra-low-field magnetodielectric effect of the $Bi_6Fe_2Ti_3O_{18}$ film is understood by coexistence of spin order and ferroelectric activity at Fe sites and the influence of low-dimension spin phase coherence under the weak magnetic field. Our findings are expected applicable on low-cost detectors and switches for very low magnetic fields in an acceptable temperature range 55K-190K.

## Acknowledgements

Authors wish to appreciate Dr. Guanghua Yu and Prof. Yue Zhang extensively for offering facilities to deposit platinum electrodes.

## References

[1]   Schmid H 2000 *Proc. SPIE* 4097 12-24
[2]   Goto T, Kimura T, Lawes G, Ramirez A P and Tokura Y 2004 *Phys. Rev. Lett.* 92 257201-04
[3]   Hemberger J, Lunkenheimer P, Fichtl R, Krug von Nidda H A, Tsurkan V and Loidl A 2005 *Nature* 434 364-7
[4]   Hur N, Park S, Guha S, Borissov A, Kiryukhin V and Cheong S W 2005 *Appl. Phys. Lett.* 87 042901-3
[5]   Kaiju H, Fujita S, Morozumi T and Shiiki K 2002 *J. Appl. Phys.* 91 7430-2
[6]   Singh M P, Prellier W, Mechin L and Raveau B 2006 *Appl. Phys. Lett.* 88 012903-5
[7]   Prasad N V and Kumar G. S 2000 *J. Magn. Magn. Mater.* 213 349-56
[8]   Suryanarayana S V, Srinivas A and Singh R S 1999 *Proc. SPIE* 3903 232-65
[9]   Cummins S E and Cross L E 1968 *J. Appl. Phys.* 39 2268-74
[10]  Kubel F and Schmid H 1992 *Ferroelectrics* 129 101-12
[11]  Pierre A C 1998 *Introduction to sol-gel processing* (Boston: Kluwer Academic Publishers)
[12]  Tahan D, Safari A and Klein L C 1994 *IEEE Int. Symp. on Applications of Ferroelectrics* (University Park: IEEE) p 427
[13]  Fuierer P and Li B 2002 *J. Am. Ceram. Soc.* 85 299-304
[14]  Liu Y L, Zhang K L, Wang F and Han Y P 2003 *Microelectron. Eng.* 66 433-437
[15]  Sporn D, Merklein S, Grond W, Seifert S, Wahl S and Berger A 1995 *Microelectron. Eng.* 29 161-8
[16]  Du X F and Chen I W 1998 *J. Am. Ceram. Soc.* 81 3253-9
[17]  Sawyer C B and Tower C H 1930 *Phys. Rev.* 35 269-75
[18]  Kim S K, Miyayama M and Yanagida H 1996 *Mater. Res. Bull.* 31 121-31

[19] Biskup N, de Andres A, Martinez J L and Perca C 2005 *Phys. Rev. B* 72 024115-21
[20] Schiller R and Nolting W 2001 *Phys. Rev. Lett.* 86 3847-50
[21] Falqui A, Lampis N, Alessandra G L and Pinna G 2005 *J. Phys. Chem. B* 109 22967-70
[22] Kurobe A, Takemori T and Kamimura H 1984 *Phys. Rev. B* 52 1457-60
[23] Saito M, Higashinaka R and Maeno Y 2005 *Phys. Rev. B* 72 14422-6
[24] Hur N, Park S, Sharma P A, Guha S and Cheong S W 2004 *Phys. Rev. Lett.* 93 107207-10
[25] Hikami S, Larkin A I and Nagaoka Y 1980 *Prog. Theor. Phys.* 63 707-10
[26] Rammer J 1991 Rev. Mod. Phys. 63 781-817
[27] Ishida S, Takeda K, Okamoto A and Shibasaki I 2004 *Physica E* 20 211-15
[28] Wang Y X, Zhong W L, Wang C L and Zhang P L 2001 *Phys. Lett. A* 288 45-8

List of figures:

**Figure 1.** Schematic illustration of sol processing.

**Figure 2.** Morphology and structure of $Bi_6Fe_2Ti_3O_{18}$ film in a macro view (a), low magnification under optical microscopy (b), and high magnification under SEM (c).

**Figure 3.** XRD patterns of $Bi_6Fe_2Ti_3O_{18}$ film (a), $Bi_6Fe_2Ti_3O_{18}$ powder derived from the same sol as film (b), and $Bi_4Ti_3O_{12}$ powder from JCPDS, which is single phase(c).

**Figure 4.** Dielectric constants (a) and relative variance of dielectric constants in magnetic field of 10Oe, i.e., $[\varepsilon_r(H)-\varepsilon_r(0)]/\varepsilon_r(0)$ (b) of $Bi_6Fe_2Ti_3O_{18}$ film versus temperature. Open triangle denotes cooling without field, dotted line denotes warming without field and fine solid line warming in magnetic field of 10 Oe.

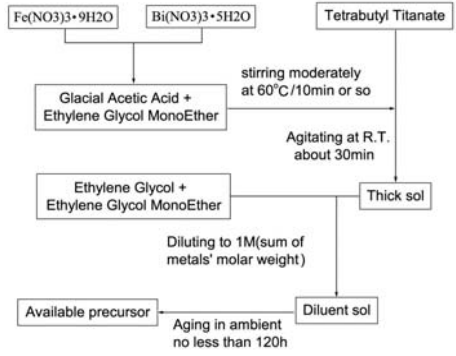

**Figure 1.** Schematic illustration of sol processing.

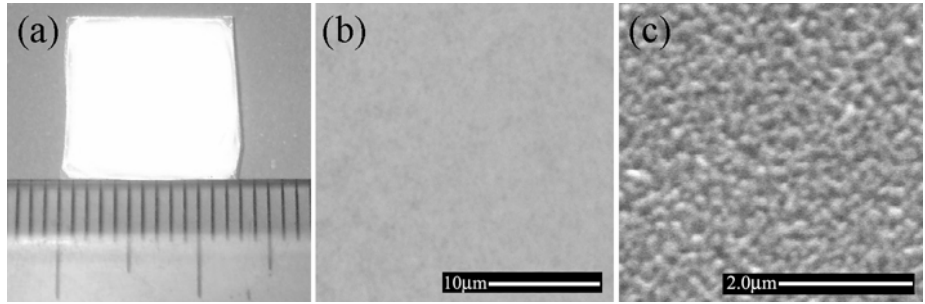

**Figure 2.** Morphology and structure of $Bi_6Fe_2Ti_3O_{18}$ film in a macro view (a), low magnification under optical microscopy (b), and high magnification under SEM (c).

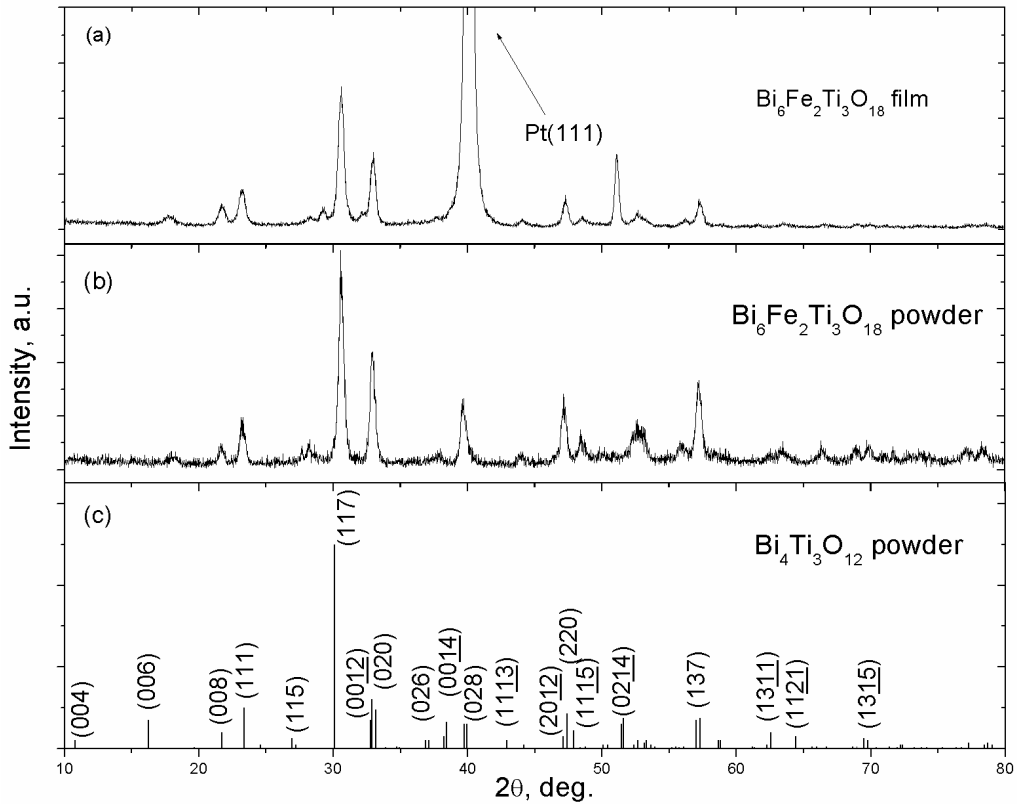

**Figure 3.** XRD patterns of $Bi_6Fe_2Ti_3O_{18}$ film (a), $Bi_6Fe_2Ti_3O_{18}$ powder derived from the same sol as film (b), and $Bi_4Ti_3O_{12}$ powder from JCPDS, which is single phase(c).

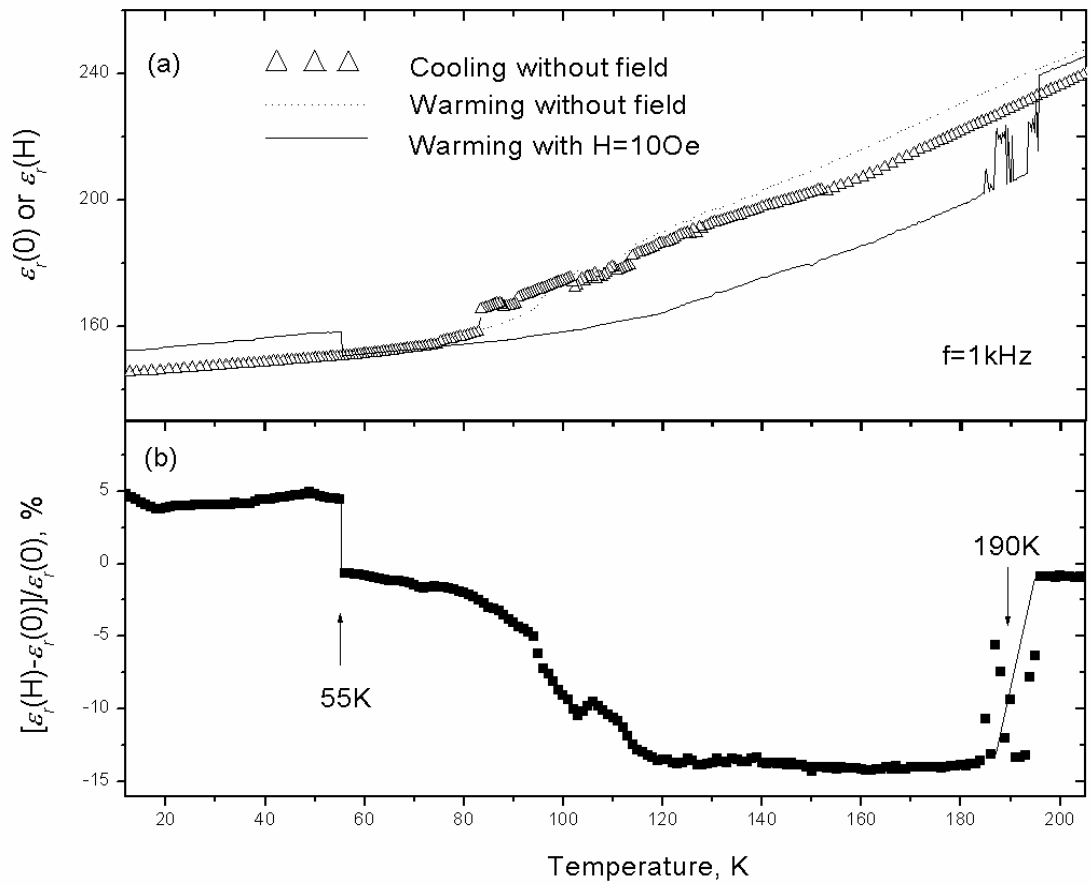

**Figure 4.** Dielectric constants (a) and relative variance of dielectric constants in magnetic field of 1OOe, i.e., $[\varepsilon_r(H)-\varepsilon_r(0)]/\varepsilon_r(0)$ (b) of $Bi_6Fe_2Ti_3O_{18}$ film versus temperature. Open triangle denotes cooling without field, dotted line denotes warming without field and fine solid line warming in magnetic field of 10 Oe.